 \documentstyle[12pt]{article}
 
\newcommand{\be}{\begin{equation}}
\newcommand{\ee}{\end{equation}}
\newcommand{\bea}{\begin{eqnarray}}
\newcommand{\eea}{\end{eqnarray}}
\newcommand{\beann}{\begin{eqnarray*}}
\newcommand{\eeann}{\end{eqnarray*}}
\newcommand{\beasn}{\begin{sneqnarray}}
\newcommand{\eeasn}{\end{sneqnarray}}

\catcode`@=11
\def\dif{{\rm d}}
\def\deriv{\@ifnextchar[{\@deriv}{\@deriv[]}}
   \def\@deriv[#1]#2#3{\mathchoice%
{{\dif^{#1}#2\over\dif{#3}^{#1}}}{{\dif^{#1}#2/\dif{#3}^{#1}}}%
{{\dif^{#1}#2\over\dif{#3}^{#1}}}{{\dif^{#1}#2/\dif{#3}^{#1}}}}

\def\presup#1{{}^{#1}\kern-.15em\relax}      %pre-superscript
\def\presub#1{{}_{#1}\kern-.12em\relax}      %pre-subscript

\def\secteqno{\@addtoreset{equation}{section}%
\def\theequation{\thesection.\arabic{equation}}}
\def\endsecteqno{\def\theequation{\@ifundefined{chapter}%
{\arabic{equation}}{\thechapter.\arabic{equation}}}}
\newcounter{subequation}
\def\thesubequation{\alph{subequation}}
\def\sneqnarray{\stepcounter{equation}\let\@currentlabel=\theequation
\setcounter{subequation}{1}
\def\@eqnnum{{\rm (\theequation\thesubequation)}}
\global\@eqcnt\z@\tabskip\@centering\let\\=\@eqncr\let\@@eqncr=\@@sneqncr
$$\halign to \displaywidth\bgroup\@eqnsel\hskip\@centering
 $\displaystyle\tabskip\z@{##}$&\global\@eqcnt\@ne
 \hskip 2\arraycolsep \hfil${##}$\hfil
 &\global\@eqcnt\tw@ \hskip 2\arraycolsep $\displaystyle\tabskip\z@{##}$\hfil
  \tabskip\@centering&\llap{##}\tabskip\z@\cr}
\def\endsneqnarray{\@@sneqncr\egroup $$\global\@ignoretrue}
\def\@@sneqncr{\let\@tempa\relax
   \ifcase\@eqcnt \def\@tempa{& & &}\or \def\@tempa{& &}
   \else \def\@tempa{&}\fi
     \@tempa \if@eqnsw\@eqnnum\stepcounter{subequation}\fi
     \global\@eqnswtrue\global\@eqcnt\z@\cr}
\def\nobiblabels{\def\@lbibitem[##1]##2{\@bibitem{##2}}}
\catcode`@=12
\secteqno
\begin{document}

%%%%%%%%%%%%%%%%%%%%%%%%TITLE%%%%%%%%%%%%%%%%%%%%%%%%%%%%%%%%%%%%%%%%%%%%

\title{{\bf Heavy Quarkonium and
            \\ non-perturbative corrections}}

\author{{\sc A.\,Pineda}
        {\sc and J.\,Soto}\\
        \small{\it{Departament d'Estructura i Constituents
               de la Mat\`eria}}\\
        \small{\it{and}}\\
        \small{\it{Institut de F\'\i sica d'Altes Energies}}\\
        \small{\it{Universitat de Barcelona}}\\
        \small{\it{Diagonal, 647}}\\
        \small{\it{E-08028 Barcelona, Catalonia, Spain.}}\\
        {\it e-mails:} \small{pineda@ecm.ub.es, soto@ecm.ub.es} }

\date{\today}

\maketitle

\thispagestyle{empty}

\begin{abstract}

We analyse the possible existence of non-perturbative contributions in
heavy $\bar Q Q$ systems ($\bar Q$ and $Q$ need not have the same
flavour) which cannot be expressed in terms of local condensates.
Starting from QCD, with well defined approximations and splitting
properly the fields into large and small momentum components, we derive
an effective lagrangian where hard gluons (in the non-relativistic
aproximation) have been integrated
out.
The
large momentum contributions (which are dominant) are calculated using
Coulomb
type states.
Besides the usual condensate corrections, we see the possibility of
new non-perturbative contributions. We parametrize them in terms of two
low momentum correlators
with Coulomb bound state energy insertions $E_n$.
We realize that the Heavy Quark Effective lagrangian can be used in
these correlators.
 We
calculate the corrections that they give rise to in the decay
constant, the bound state energy and the matrix elements of bilinear
currents at zero recoil. We study the cut-off dependence of the new
contributions and we see that it matches perfectly with that of the
large momentum contributions.
 We consider two situations in detail:
i) $E_n>> \Lambda_{QCD}$ ($M_Q \rightarrow \infty$) and ii) $E_n <<
\Lambda_{QCD}$, and briefly discuss the expected size of the new
contributions in $\Upsilon$ , $J/\Psi$ and $B_{c}^{\ast}$ systems.

\end{abstract}
\bigskip
PACS: 12.38.Lg, 12.39.Hg, 13.20.Gd

\vfill
\vbox{
\hfill February 1996\null\par
\hfill UB-ECM-PF 96/03}\null\par

\clearpage

%%%%%%%%%%%%%%%%%%%%%%%%%%%%%%%%%%%%%%%%%%%%%%%%%%%%%%%%%%%%%%%%%%%%%%%%%

%%%%%%%%%%%%%%%%%%%%%%%INTRODUCTION%%%%%%%%%%%%%%%%%%%%%%%%%%%%%%%%%%%%%%%

\section{Introduction}
\indent

\def\np{Nucl.Phys. }\def\pl{Phys.Lett. }\def\ap{Ann.Phys. }
\def\pr{Phys.Rev. }\def\prl{Phys.Rev.Lett. }\def\prp{Phys.Repts. }
\def\ijmp{Int.J.Mod.Phys. }\def\mpl{Mod.Phys.Lett. }\def\jp{J.Phys. }
\def\rmp{Rev.Mod.Phys. }

\def\a{\alpha}  \def\b{\beta} \def\g{\gamma} \def\G{\Gamma}
\def\z{\zeta} \def\th{\theta} \def\TH{\Theta} \def\tth{\vartheta}
\def\k{\kappa} \def\l{\lambda} \def\L{\Lambda} \def\m{\mu} \def\n{\nu}
\def\cs{\xi} \def\Cs{\Xi} \def\p{\pi} \def\P{\Pi} \def\r{\rho} \def\s{\sigma}
\def\S{\Sigma} \def\t{\tau} \def\y{\upsilon} \def\Y{\upsilon}
\def\f{\phi} \def\F{\Phi} \def\x{\chi} \def\ps{\psi} \def\Ps{\Psi}
\def\o{\omega} \def\O{\Omega} \def\vf{\varphi}
\def\pa{\partial} \def\da{\dagger} \def\dda{\ddagger}

\def \v {\rlap\slash{v}}
\def \k {\rlap\slash{ \rm k}}
\def \el {\rlap\slash{\rm e}}
\def \DEL {\rlap\slash{\Delta}}
\def \Qp {Q^{\prime}}
\def \Qpp {Q^{\prime\prime}}
\def \kp {k^{\prime}}

\bigskip

The study of heavy quark bound state systems remains one
of the more promising topics in order to test both perturbative and
non-perturbative aspects of QCD, as it is clear from the steady
activity in the field \cite{nuevonrqcd}-\cite{nos1}. These systems
can be understood in a first approximation as non-relativistic bound
states which occur due to a
Coulomb type interaction predicted by perturbative QCD.
In order to improve this basic picture one has to deal on one side with
perturbative relativistic and radiative corrections, and on the other
side with non-perturbative
corrections (power corrections).

\medskip

In this paper we shall only be concerned with non-perturbative
corrections. Usually, the latter have been parametrized
using both the multipole expansion and the adiabatic approximation in
terms of
the gluon condensate \cite{Vol,Leut}. Corrections to the
Coulomb potential due to condensates can also be considered,
although these are subleading \cite{Ynd,subpotential}. We have argued
before \cite{nos1} that new non-perturbative contributions could arise
 which cannot be expressed
in terms of local condensates, and hence a convenient parametrization
for them is required. This kind of nonperturbative contributions has
been discussed
in \cite{Shiftman} in a different context and, in fact, the various
Isgur-Wise functions
extensively used in the Heavy Quark Effective Theory (HQET) may be
regarded as such \cite{IW}.

\medskip

Let us recall the main idea behind the possibility of new
non-perturbative contributions in heavy quarkonium\footnote{We
use 'heavy quarkonium' to denote a general heavy quark-antiquark bound
state. The quark and the antiquark need not have the same flavour.}.
 When the relative
three momentum in the
bound state is big enough
the dominant
interaction must be the perturbative Coulomb potential,
 but for small relative three momentum this need not
longer be true. Therefore, heavy quarks in the latter kinematical
situation should
better be kept as low energy degrees of freedom. It turns out that a
convenient parametrization of this kinematical region may be given in
terms of the HQET for quarks and antiquarks \cite{nos1,joan2}.

\medskip

The
HQET for quarks
and antiquarks enjoys rather peculiar features, which make it quite
different from the usual HQET describing either quarks or antiquarks,
which has been so popular in the study of $Q\bar q$ and $Qqq$ systems in recent
years \cite{hqet} (see \cite{reviews} for reviews).
 For instance, it enjoys a symmetry, which is larger
than the
well-known spin and flavor symmetry, that breaks spontanously down to
the latter giving rise to quark antiquark states as Goldstone modes
\cite{joan2}. Its pecularities concerning radiative corrections have
recently been illustrated in \cite{nos2}.

\medskip

The main aim of this paper is to work out a controlled derivation from
QCD of the effective
lagrangian describing the small relative momentum regime of heavy quarks
in quarkonium. Whereas the basic ideas above have
already
been elaborated in \cite{nos1}, a complete and systematic
derivation is lacking, and hence worth being presented.
 Within this new framework we
recalculate the non-perturbative contributions of
this region to the energy levels, the decay constant and the matrix
elements
of bilinear currents at zero recoil. We find a few corrections to the
formulas given in \cite{nos1}. For all these observables it is
enough to work
in the center of mass frame (CM), which we shall do in most of the
paper.

\medskip

In order to deal with heavy quarkonia systems we keep
the relevant degrees of freedom in the QCD lagrangian.
In fact, since virtual heavy quark creation is very much suppressed, we
could safely start from non-relativistic QCD (NRQCD). The derivation of
NRQCD from QCD is well understood and a technique to incorporate
relativistic corrections  to it has also been developed \cite{nrqcd}.
First of all,
we split the gluon field in hard and soft by a three momentum cut-off.
From the hard gluon fields we only keep the zero component and
disregard the spacial components.
This is legitimate as far as we are not interested in
relativistic corrections.
We next integrate out the zero component of the hard gluon field to
obtain the Coulomb potential between heavy quark currents. The Coulomb
potential has an infrared momentum cut-off since the zero component of
the soft gluon field has not been integrated out.
At this point we have an effective lagrangian formally equal to the one
used by Voloshin and Leutwyler (VL) \cite{Vol,Leut}, except for the IR
cut-off in the Coulomb potential.
After introducing CM and relative momentum for the
bound states we are interested in, we further split the quark fields in
large and small relative three-momentum
regimes\footnote{The large and small relative momentum regions were
denoted as off- and on-shell regions respectively in \cite{nos1}.}.
 The resulting lagrangian can then be separated in three
pieces:
$L^{\mu}$
 which contains small relative momentum quark fields
only,
$L_{\mu}+L^{I}_{\mu}$
contains large relative momentum quark fields only and
$L^{I\mu}_{\mu}$
which contains
both small and large relative momentum quark fields. For
$L^{\mu}$
 we can aproximate the lagrangian to
the HQET lagrangian, where eventually all its powerful symmetries
can be used. No Coulomb term remains in this part of the lagrangian. For
$L_{\mu}+L^{I}_{\mu}$
we obtain again the VL starting point lagrangian except for two facts:
both the Coulomb potential and the
Hilbert space are restricted
to three-momenta larger than a certain cut-off. Keeping the
cut-off much
higher than $\Lambda_{QCD}$ but much smaller than the invers Bohr
radius we may safely assume that the
multipole expansion holds for this part of the lagrangian. If we further
assume that the adiabatic approximation also holds, we may
proceed in total analogy to VL. The hipothesis above on the
cut-off also allows us to
treat
$L^{I\mu}_{\mu}$
 as a perturbation. The various contributions from this
perturbation to the different observables can be eventually expressed as
correlators of the HQET.

\medskip

We would like to stress that our formalism is less restrictive than
the one used by VL since neither the adiabatic approximation nor the
multipole expansion are assumed to hold in the small relative momentum
region of the heavy quark fields. Indeed we may always recover VLs
results by putting to zero the cut-off which separates large and small
relative momentum.

\medskip

We distribute the paper as follows. In sect. 2 we derive the
effective action for the small relative momentum fields.
 In sect. 3 we calculate the decay constant, the bound state mass and
the matrix
element of any bilinear heavy quark current between quarkonia states at
zero recoil.
The latter is relevant for the study of
semileptonic
decays at zero recoil. In sect. 4 we prove the cut-off independence of
our results. In section 5 we study the low momentum correlators
in two situations:
 the asymptotic limit ($M_{Q} \rightarrow \infty$) $E_{n} >>
\Lambda_{QCD}$, where, using OPE techniques,
 we see that no
new corrections arise,
and ii)  $E_{n} << \Lambda_{QCD}$, where
 the low momentum contributions are evaluated using an
effective 'chiral' lagrangian  which incorporates the relevant
symmetries of the HQET for quarks and antiquarks.
Working in this way we find new
non-perturbative contributions which are parametrized by a single
non-perturbative constant. We also give preliminary estimations of
their size. Section 6 is devoted to the conclusions.

%%%%%%%%%%%%%%%%%%%%%%%EFFECTIVE ACTION%%%%%%%%%%%%%%%%%%%%%%%%%%%%%%%%%

\def\p{\prime}
\def\qp{q^{\prime}}
\def\ap{a^{\prime}}
\def\kp{k^{\prime}}
\def\vap{\vec q^{\prime}}
\def\vkp{\vec k^{\prime}}
\def\vq{\vec q}
\def\vk{\vec k}

\section{Effective action}
\indent

In this section we derive the effective lagrangian
for heavy quarks and antiquarks in the small relative momentum regime
from QCD.

The QCD lagrangian reads
\be
{\cal L}= -{1\over4}F^{2}+ \sum_{a}{\bar
Q_{a}}(i{D\!\!\!\!/}-m_{a})Q_{a}
\ee
where
\be
D_{\mu}= \partial_{\mu}-igV_{\mu}\,, \quad V=V^{r}T^{r}
\ee
\be
F^r_{\mu\nu}=\partial_{\mu}V^r_{\nu}-\partial_{\nu}V^r_{\mu}+gf^{rst}
V_{\mu}^{s} V_{\nu}^{t}
\ee

We split the gluon field $V$ in large $A$ and small $B$
momentum modes
% through a hard three momentum cut-off $\mu$.
%\be
%\label{gluon}
$V(x)=A(x)+B(x)$.
%\,, \quad
%V(x)= \int {d^3k\over(2\pi)^3} e^{i {\vec k}.{\vec x}} {\tilde V}(\vec
%k,t)\,,
%\ee
%$$
%\quad {\tilde A}(\vec k,t)= {\tilde V}(\vec k,t) \theta (k-\mu)\,,
%\quad {\tilde B}(\vec k,t)= {\tilde V}(\vec k,t) \theta (\mu -k)\,.
%$$
Next we exactly integrate $A_0$ and neglect $A_{i}$. The latter would
give rise to relativistic corrections. Consistently, at the same point
we perform a Foldy-Wouthuysen transformation and keep terms up to $1/m$.
We obtain
\bea
\nonumber
&&{\rm L}=-{1\over4}\int d^3x F^{2}_{B}+ \sum_{a} \int d^3x \left(
{\bar Q_{a}}(i \gamma^{0} D_{0}^{B}-m_{a})Q_{a} +
{\bar Q_{a}}{{\vec {D}}_{B}^2\over 2m_{a}}Q_{a}+
{\bar Q_{a}}{g \vec \Sigma.\vec B_{B}\over
2m_{a}}Q_{a} \right)
\\
&& + O({1\over m_{a}^2})
-{g^2\over2}
\sum_{aa^{\prime}}
\int d^3 x
\int d^3 y
\bar Q_{a}\gamma^0T^{r}Q_{a}(x) \left(  {1\over {\vec
D_{B}}^{2}}\right)^{rs}(x,y)
\bar Q_{a^{\prime}}\gamma^0T^{s}Q_{a^{\prime}}(y)
\eea
%\end{document}
%Although it is not essential, we will fix the gauge as a matter of
%convenience.
%We use the Coulomb gauge fixing in momentum space for the large
%momentum modes.
%Other similar gauges can be used without essential differences (for
%instance, the background Coulomb gauge fixing), which only arise when
%relativistic corrections are calculated.
% \be
%k_{i} {\tilde A_{i}}(\vec k,t)=0
%\ee
which is manifestly gauge invariant
\footnote{Similar
approaches can be found in the literature \cite{Nunes}.}
. Although, in principle, we could
attempt to carry out an explicitely gauge invariant calculation, in
practise, it is most convenient to choose
 a slightly modified Schwinger gauge for the small momentum gluons
\be
(\vec z -{m_{a}\vec x +m_{a^{\prime}}\vec y\over
m_{a}+m_{a^{\prime}}})\vec B (z)=0 \ee
In this gauge  $\vec B$ in the kinetic
and Coulomb terms gives rise to subdominant contributions when the
multipole expansion is carried out,
 % in order to keep the time dependence explicit.
%\be
%{\vec \nabla_{k}} {\vec{\tilde B}}(\vec k,t)=0
%\ee
which greatly simplifies the calculation. In particular, recall that the
propagator in the Coulomb term always carries large momentum (we have
not integrated
out the small momentum $V_0$ which is kept in $B_0$). Hence the
multipole expansion is always legitimated in the Coulomb term. This
allows to drop $\vec B$ in the Coulomb term straight away.
% arising from
%the multipole expansion. % Next we exactly integrate $A_{0}$ and
%neglect %$A_{i}$, since the latter
% would give rise to relativistic corrections.
%At this point we perform a
%Foldy-Wouthuysen transformation. The four-fermion interaction term
%remains unchanged.
As long as we are interested in quark-antiquark
 bound states only, we may also safely neglect
the four-fermion interaction terms involving only quarks or only
antiquarks.
We next rearrange the quark-antiquark interaction term in a
convenient way
 in order to describe the bound state dynamics
.
Finally, the effective lagrangian reads
\bea
\nonumber
&&{\rm L}=-{1\over4}\int d^3x F^{2}_{B}+ \sum_{a} \int d^3x \left(
{\bar Q_{a}}(i \gamma^{0} D_{0}^{B}-m_{a})Q_{a} +
{\bar Q_{a}}{{\vec D_{B}}^2\over 2m_{a}}Q_{a}
%{\bar Q_{a}}{g \vec \Sigma.\vec B_{B}\over
%2m_{a}}Q_{a}
\right)
\\
&& + O({1\over m_{a}^2})
-{1\over2}
\sum_{aa^{\prime}A}m^3_{aa^{\prime}}N^2_A \sum_s
\int {d^3v\over(2\pi)^3}
\int {d^3q^{\prime}\over(2\pi)^3} \int {d^3q\over(2\pi)^3}
V^A
(\vec q^{\prime}- \vec q)
\\
\nonumber
&& \times \left[ \tilde {\bar Q}_{a}(-m_a \vec v+\vec
q,t)
T^A \bar \Gamma_{s}
\tilde Q_{a^{\p}}(m_{a^{\prime}} \vec v+\vec
q,t) \right]
\left [\tilde {\bar Q}_{{\ap}}(m_{\ap} \vec v+\vec
\qp,t)
T^A \Gamma_{s}
\tilde Q_{a}(-m_a \vec v+\vec
\qp,t) \right]
\eea
where $A={0,r}$ denotes colour ($0$ singlet and $r$ octet, $r=1...
8$), $|\vec q- \vec q^{\prime}| > \mu$ , $\mu$ being the cut-off which
separates small and large momenta, and
\be m_{a\ap}=m_a+m_{\ap}\,, \quad \Gamma_{s}=i\gamma_5p_{-}\,,
i\gamma^{i}p_{-}, \quad p_{\pm}:={1\pm \gamma^0 \over 2}\,,
\ee
$$
N_A={1\over \sqrt{N_c}},\sqrt2 \,, \quad
T^A=1, T^r \,.
$$
While, the potential reads
\be
\label{pot}
V^0(\vec p)= -{C_{F} g^2 \over {\vec p}^2}\,, \quad
V^r(\vec p)= {g^2 \over 2N_c{\vec p}^2}\,.
\ee
where $C_{F}=(N_{c}^2-1)/2N_{c}$ and
$\vert \vec p\vert >\mu$ must be understood due to
 the cut-off coming from soft gluons{\footnote{Several aspects related
to this cut-off dependence have been studied in \cite{nos2}.}}.

\medskip

Written in this way, we can understand the four-fermion Coulomb
interaction term as one which creates a quark-antiquark state
with central velocity ${\vec v}$ and relative momentum ${\vec
\qp}$ and annihilates a quark-antiquark state with the same center of masses
velocity ${\vec v}$ and relative momentum ${\vq}$. Obviously ${\vec v}$
is a conserved quantity in this
non-relativistic approximation. We consider the spin breaking term as
subleading and we will neglect it in the following. Therefore, spin
symmetry for both low and high momentum is implicit in the rest of the
paper.

\medskip

If we stopped at this point we would obtain the standard VL results.
However, we would like to go beyond and look for
new non-perturbative contributions. We observe that
quarks with small relative three momentum only feel the
Coulomb
interaction of quarks with large relative momentum.
This suggests to perform a splitting of the physical
quark and antiquark fields
into small and large relative momentum in the
bound state. The physical picture behind  is that if the
relative three
momentum in the bound state is big enough we can understand it as a
perturbative Coulomb type bound state. But for small relative three
momentum that is no longer true. For that momentum regime the quark and
antiquark fields should be kept as low momentum degrees of freedom. That
is, in fact, the main idea of the paper.
Therefore, let us write down the currents related to the physical
quark-antiquark states in momentum space
\bea
\label{phcur}
\nonumber
&&J_{\Gamma}^{A,\ap a}(x)= \bar Q_{\ap} T^A \Gamma Q_a (x)=
\\ &&
m^3_{a\ap} \int {d^3v\over(2\pi)^3} e^{im_{a\ap}\vec v.\vec x}
\int {d^3q\over(2\pi)^3}
\tilde {\bar Q}_{\ap}(-m_{\ap} \vec v-\vec
q,t)
T^A \Gamma
\tilde Q_{a}(m_a \vec v-\vec
q,t)\,.
\eea
The matrix $\Gamma$ should be such that it projects over quark-antiquark
states according to our non-relativistic picture. Notice that
 the time dependence is kept explicit.
Furthermore, we split the relative
three momentum with the same cut-off $\mu$ as
above. Thus, (\ref{phcur}) reads
\bea
\nonumber
&&J_{\Gamma}^{A,\ap a}(x)=J_{l,\Gamma}^{A,\ap a}(x) + J_{h,
\Gamma}^{A,\ap a}(x)
\\
\nonumber
&=& m^3_{a\ap} \int {d^3v\over(2\pi)^3} e^{im_{a\ap}\vec v.\vec x}
\int^{\mu} {d^3q\over(2\pi)^3}
\tilde {\bar h}^v_{\ap}(-\vec
q,t)
T^A \Gamma
\tilde h^v_{a}(-\vec
q,t)
\\
&+& m^3_{a\ap} \int {d^3v\over(2\pi)^3} e^{im_{a\ap}\vec v.\vec x}
\int_{\mu} {d^3q\over(2\pi)^3}
\tilde {\bar Q}_{\ap}(-m_{\ap} \vec v-\vec
q,t)
T^A \Gamma
\tilde Q_{a}(m_a \vec v-\vec
q,t)\,,
\eea
where $\tilde Q_a(m_a \vec v +\vk)=:
\tilde h^v_a (\vk)
$.
After that we may divide the lagrangian
in three pieces.
\be
L= L_{\mu}+L^{\mu}+L^I
\ee
\medskip
$L_{\mu}$ is the piece of the effective lagrangian containing large
momenta only. It reads
\bea
\nonumber
&&{\rm L_{\mu}}= \sum_{a} \int d^3x \left( {\bar
Q_{a}}(i \gamma^{0} \partial_{0}-m_{a})Q_{a} +
{\bar Q_{a}}{{\vec \nabla}^2\over 2m_{a}}Q_{a} \right)
\\
&&
 -{1\over2}
\sum_{aa^{\prime}A}m^3_{aa^{\prime}}N^2_A \sum_s
\int {d^3v\over(2\pi)^3}
\int_{\mu} {d^3q^{\prime}\over(2\pi)^3} \int_{\mu} {d^3q\over(2\pi)^3}
V^A(\vec q^{\prime}- \vec q)
\\
\nonumber
&& \times \left[ \tilde {\bar Q}_{a}(-m_a \vec v+\vec
q,t)
T^A \bar \Gamma_{s}
\tilde Q_{a^{\p}}(m_{a^{\prime}} \vec v+\vec
q,t) \right]
\left[ \tilde {\bar Q}_{{\ap}}(m_{\ap} \vec v+\vec
\qp,t)
T^A \Gamma_{s}
\tilde Q_{a}(-m_a \vec v+\vec
\qp,t) \right]
\eea
where $|\vec q- \vec q^{\prime}| >
\mu$.

In fact it is nothing but the standard Coulomb lagrangian,
except for the cut-offs.
$L^{\mu}$ is the piece of the effective lagrangian containing small
momenta only. It reads

\be
\label{Lmu}
{\cal L^{\mu}}= -{1\over4}F^{2}_{B}+ \sum_{a}\left\{ {\bar
h^v_{a}}(i \gamma^{0} D_{0}^{B}-m_{a})h^v_{a} +
{\bar h^v_{a}}{{\vec {D}}_{B}^2\over 2m_{a}}h^v_{a} \right\}\,.
\ee
Notice that (\ref{Lmu}) does not have the four-fermion
Coulomb
term. It contains the whole soft gluon lagrangian as well as the heavy
quark and antiquark fields
with small three relative momentum.
 All the fields in
(\ref{Lmu})
are in the
non-perturbative regime of QCD.
Notice that if we drop the term in $1/m_{a}$ and make
$
 h_a^v \rightarrow e^{-i\gamma^0m_{a} x_0}h_a^v
$
(\ref{Lmu}) becomes the HQET lagrangian in the rest frame.
Although the $1/m_a$ term is naively subleading for small relative
momentum, it plays a crucial role in certain circumstances, as we
shall see in Section 4. Nonetheless let us advance that for the
correlators we will be interested in one can safely neglect it
 and work with the HQET lagrangian.
\medskip

$L^I$ mixes small and large momenta
\be
L^I=L^I_{\mu}+L^{I\mu}_{\mu}
\ee
The first term reads

\be
\label{lagvol}
{\cal L}^I_{\mu}(x)= g{\bar Q}_a\gamma^0 B^r_0 T^r Q_a(x)
\ee
which
 gives the leading contribution to the multipole
expansion.
We will not discuss these contributions (\ref{lagvol}) here
since they
have been extensively studied in the literature \cite{Ynd,Vol,Leut}.
Let us focus on the second term.
 It reads
\bea
\label{lagint}
&&L^{I\mu}_{\mu}=-{1\over2} \sum_{aa^{\prime}A}m^3_{aa^{\prime}}N_A^2
\sum_s
\int {d^3v\over(2\pi)^3}
\int_{\mu}{d^3q^{\prime}\over(2\pi)^3} \int^{\mu} {d^3q\over(2\pi)^3}
V^A(\vec q^{\prime}- \vec q)
\\
\nonumber
&& \times \left[ \tilde {\bar h^v}_{a}(\vec
q,t)
T^A \bar \Gamma_{s}
\tilde h^v_{a^{\p}}(\vec
q,t) \right]
\left[ \tilde {\bar Q}_{{\ap}}(m_{\ap} \vec v+\vec
\qp,t)
T^A \Gamma_{s}
\tilde Q_{a}(-m_a \vec v+\vec
\qp,t) \right] + (h.c.)\,.
\eea
In this expression the Coulomb potential is the only piece which mixes
small and large relative momentum. We can perform a derivative
expansion since $q$ and $\qp$ belong to different momentum regimes
($q \sim \Lambda_{QCD} << \qp \sim {m \a \over n}$) and keep
only the leading term (further orders would give subleading
corrections).
 It turns out that the small relative momentum term decouples from the
Coulomb potential and can be written like a local current.
 Finally, we obtain

\bea
&&L^{I\mu}_{\mu}=
-{1\over2}
\sum_{aa^{\prime}A}m^3_{aa^{\prime}}N^2_A
\sum_s
\int {d^3v\over(2\pi)^3}
\int_{\mu}{d^3q \over(2\pi)^3}
V^A (\vec q)
\\
\nonumber
&& \times \int d^3x e^{-im_{a\ap}\vec v. \vec x} J_{l,\bar
\Gamma_s}^{A,a\ap}(\vec
x,t) \left[ \tilde {\bar Q}_{\ap}(m_{\ap} \vec v+\vec
q,t)
T^A \Gamma_{s}
\tilde Q_{a}(-m_a \vec v+\vec
q,t) \right] + (h.c.)\,.
\eea

The formalism developed in
 \cite{nos1} was not powerful enough as to uncover the interaction
lagrangian (2.17).
This interaction lagrangian is indeed the responsible for the
differences between
the results presented there and the ones obtained in the next section.

\medskip

If we assume that small momentum terms are small in comparison with
the large momentum terms we can treat
the interaction lagrangian (2.17) as a perturbation.
This is so
for the lower energy levels
of heavy quark bound states.
 In the next
sections we focus on the nonperturbative contributions coming from
(2.13) and (2.17).

%%%%%%%%%%%%%%%%%%%%%%%%%%%%%%%%%%%%%%%%%%%%%%%%%%%%%%%%%%%%%%%%%%%%%%%%%%%%%%
\section{Physical Observables}
\indent

In this section we work out the non-perturbative corrections from the
small relative momentum region to the decay
constant, the bound state mass and the matrix elements of bilinear
currents at zero recoil. We take the bound state velocity small or zero.

Consider first the eigenvalues and eigenstates of
 $H_{\mu}$,
 the hamiltonian associated to
 $L_{\mu}$.
 They read
\be
\label{nstate}
\vert (ab, n, s, A);
\vec v
\rangle =
{N_A \over \sqrt{2} }
\int_{\mu} {d^3k \over (2\pi)^3}
\tilde \Phi^A_{ab,n}(\vec k;\mu)
\bar u^{\alpha} (\vec p_1) \Gamma_{s}
v^{\b}(\vec p_2)
T^A_{i_1,i_2}
b^{a\dagger}_{\a,i_1}(\vec p_1)d^{b\dagger}_{\b,i_2}(\vec p_2) \vert
0\rangle\,, \ee
\be
E^A_{ab,n}(\mu)
\ee
where

\be
\vec p_1=m_{a}\vec v + \vec k\,,
\quad
\vec p_2=m_{b}\vec v - \vec k \,, \quad
\ee

$E^A_{ab,n}$,
$\Phi^A_{ab,n}(\vec x)$ and $\tilde \Phi^A_{ab,n}(\vec k)$ are
the energy, the coordinate space wave function and the momentum space
wave function of a Coulomb-type state with quantum number
$n= (n,l,m)$. $\vec v$ is the bound state 3-vector velocity. $a$ and $b$
are flavour indices and $s$ denotes spin.
 $b^{\dagger}$ and $d^{\dagger}$
are creation operators of particles and anti-particles respectively.
\be
\{
b^{a\dagger}_{\a,i_1}(\vec p_1),b^{b}_{\b,i_2}(\vec p_2) \}=(2\pi)^3
\delta^{ab}
\delta_{\a\b}
\delta_{i_1 i_2}
\delta^{3}(\vec p_1-\vec p_2)
\ee
\be
\{
d^{a\dagger}_{\a,i_1}(\vec p_1),d^{b}_{\b,i_2}(\vec p_2) \}=(2\pi)^3
\delta^{ab}
\delta_{\a\b}
\delta_{i_1 i_2}
\delta^{3}(\vec p_1-\vec p_2)
\ee
\be
\{
b^{a\dagger}_{\a,i_1}(\vec p_1),d^{b}_{\b,i_2}(\vec p_2) \}=0
\ee

$ u^{\alpha} (\vec p_1)$ and $ v^{\b}(\vec p_2) $ are spinors normalized
in such a way that in the large $m$ limit the following holds
\be
\sum_{\a} u^{\alpha} (\vec p_1) \bar u^{\alpha} (\vec p_1) =p_{+}
\,, \quad  \sum_{\a} v^{\alpha} (\vec p_1) \bar v^{\alpha} (\vec
p_1) =-p_{-} \,.
\ee

(\ref{nstate}) has the non-relativistic normalization
\be
\label{norms}
\langle  (ab,n,s,A);
\vec v
\vert
 (a^{\prime}b^{\prime},n^{\prime},s^{\prime},A^{\p});
\vec v^{\prime}
\rangle                   =
(2\pi)^3\delta^{(3)}(m_{ab}(\vec v -\vec v^{\prime}))
 \delta_{n,n^{\prime}} \delta_{s,s^{\prime}}
 \delta_{(ab),(a^{\prime}b^{\prime})}
 \delta_{A,A^{\prime}}\,,
\ee
where we have used
\be
tr\left(p_{+}\Gamma^{s}p_{-}{\bar \Gamma}^{s^{\prime}} \right) =
-2\delta_{s,s^{\prime}}
\ee

From (\ref{norms}) it follows the wave function
normalization
\be
\int_{\mu}{d^3q \over (2\pi)^3}
\tilde \Phi^{A*}_{ab,n^{\prime}}(\vec q;\mu) \tilde \Phi^A_{ab,n}(\vec
q;\mu) = \delta_{n,n^{\prime}}
\ee
where there is no sum over A.
 The wave function and the energy fulfil the equation
\be
\label{mueq}
{p^2 \over 2\mu_{ab}} \tilde \Phi^A_{ab,n}(\vec p;\mu)+\int_{\mu}{d^3q
\over (2\pi)^3} \tilde \Phi^A_{ab,n}(\vec q;\mu) V^A(\vec p - \vec q) =
E^A_{ab,n}(\mu) \tilde \Phi^A_{ab,n}(\vec p;\mu)\,,
\ee
$$
p > \mu\,, \quad \mu_{ab}={m_am_b \over m_a+m_b} \,.
$$

From (\ref{pot}) it trivially follows that
the eight components of the octet wave function fulfil the same
equation and hence they are the same.
Notice that the wave function normalization and the differential
equation above are
$\mu$ dependent. Furthermore,
 the wave function is
not defined over all values of $p$. We will work
this out in detail in sect. 4. In order to simplify the notation we will
not displayed the cut-off dependence explicitely
in the rest of the section, but it must be understood throughout.

For
$H^{\mu}$,
 the hamiltonian associated to
$L^{\mu}$,
 we denote the eigenstates
and eigenvalues
by
\be
\label{gstate}
\vert (ab, g, s);
\vec v
\rangle \,, \quad \quad
E_{g}
\ee
where $g$ labels the low momentum state.
 We cannot give explicit expressions since their dynamics
is governed by low momentum.
(\ref{gstate}) has the non-relativistic normalization
\be
\langle  (ab, g, s);
\vec v
\vert
 (a^{\prime}b^{\prime}, g^{\p}, s^{\prime});
\vec v^{\prime}
\rangle                   =
(2\pi)^3\delta^{(3)}(m_{ab}(\vec v -\vec v^{\prime}))
 \delta_{s,s^{\prime}}
 \delta_{(ab),(a^{\prime}b^{\prime})}
 \delta_{g,g^{\prime}}\,.
\ee
Of course the states (\ref{nstate})
 and (\ref{gstate})
 are orthogonal
since they belong to different momentum regimes.

Our Hilbert space is (before switching on
$L_{\mu}^{I\mu}$)
\{N\}=\{(n,A), g\} and the identity reads in this base
\be
1 \simeq \vert 0 \rangle \langle 0 \vert + 1^{\mu}+1_{\mu} =
\vert 0 \rangle \langle 0 \vert + \sum_{ab,N,s} \int{d^3
\vec P \over (2\pi )^3} \vert (ab,N,s); \vec v
\rangle \langle (ab,N,s);
\vec v
 \vert\,.
\ee

Let us now calculate the matrix elements of
$H_{\mu}^{I\mu}$,
 the hamiltonian associated to
$L_{\mu}^{I\mu}$.
 We note that the
only matrix element different from zero is
$$
\langle  (ab,g,s);
\vec v
\vert H^{I\mu}_{\mu} \vert
 (a^{\prime}b^{\prime},n,s^{\prime},A);
\vec v^{\prime}
\rangle
$$
\be
=
(2\pi)^3\delta^{(3)}(m_{ab}(\vec v -\vec v^{\prime}))
E^A_{ab,n^{\prime}} {\tilde \Phi^A_{ab,n^{\prime}}}(\vec
0){f^{A*}_{ab,g} N_A \over \sqrt2}
 \delta_{s,s^{\prime}}
 \delta_{(ab),(a^{\prime}b^{\prime})}\,,
\ee
where
\be
\langle  (ab,g,s);
\vec v \vert
{\bar h}^v_a T^A \Gamma^{s^{\p}} h^v_b(0)
\vert 0 \rangle =: f^{A*}_{ab,g} \delta_{s,s^{\p}}\,.
\ee

\bigskip
In the calculations above we have not made any explicit assumption about
the relative size of $L_{\mu}$ and $L^{\mu}$.
 We are mainly interested
in very heavy quark-antiquark bound states where small momenta can be
considered as corrections, at least for the lower energy levels.
Clearly, these bound states
should be singlets since the octet potential is repulsive.
 In fact, at the
level we are working, the octet states are not going to give
contributions to the physical observables so we will neglect them in the
following. Hence from now on colour singlets are understood and colour
indeces dropped. We also remark that we are always working in the CM
frame, even though sometimes we keep $\vec v \not= 0$ in some
intermediate steps for convenience.
 Following standard Quantum Mechanics
perturbation theory \cite{Sak} we can obtain the corrected bound state
energy\footnote{The correction to the bound state energy was found to
be zero in \cite{nos1} because the existence of $L^{I\mu}_{\mu}$ was not
known.}
 and wave functions (states) for the lower levels.
They read
\be
\label{deltam}
\delta E_{ab,n} =
{\P_{ab}(E_{ab,n}) \over 2N_c}
|E_{ab,n}{\tilde \Phi}_{ab,n}(\vec 0)|^2
\ee
\be
\label{phstate}
\vert (ab,n,s); \vec v \rangle_{F} = Z^{1/2}_{n}
\vert \overline{(ab,n,s); \vec v} \rangle_{F}
\ee
\be
\label{n}
\vert \overline{ (ab,n,s); \vec v} \rangle_{F}
= \vert (ab,n,s); \vec v \rangle +
\vert (ab,n,s); \vec v \rangle^{(1)} +
\vert (ab,n,s); \vec v \rangle^{(2)} +...
\ee
\be
\label{n1}
\vert (ab,n,s); \vec v \rangle^{(1)} =
{{\tilde \Phi}_{ab,n} (\vec 0)
E_{ab,n} \over \sqrt{2N_c}} {\hat G}^{\mu}(E_{ab,n})
J_{l,\Gamma^s}^{ab}(0) \vert 0 \rangle
\ee
\be
\label{n2}
\vert (ab,n,s); \vec v \rangle^{(2)} =
\sum_{m \not= n} \vert (ab,m,s); \vec v \rangle
{\P_{ab}(E_{ab,n}) \over 2N_c} E_{ab,n}
{\tilde \Phi}_{ab,m}^{*}(\vec 0)
{\tilde
\Phi}_{ab,n} (\vec 0) {E_{ab,m} \over
E_{ab,n}- E_{ab,m}} \ee
\be
\label{normstate}
Z_{n} \simeq 1+
{1 \over 2N_c}
{d\P_{ab}(E_{ab,n})\over dE_{ab,n}}
|E_{ab,n}{\tilde \Phi}_{ab,n}(\vec 0) |^2
\ee
where both continuum and bound states are included in the sum in
(\ref{n2}), (\ref{phstate}) denotes the physical normalized state (with
low
momentum corrections) and
\be
{\hat G}^{\mu}(z) := {1 \over z-H^{\mu}+i \epsilon }
\ee
\be
\label{cor1}
i\int d^4x e^{iP_n x} \langle 0 \vert T \left\{ J^{ba}_{l,{\bar
\Gamma}^{s^{\p}}} (x) J^{ab}_{l,\Gamma^s} (0)
\right\} \vert 0 \rangle =: \P_{ab}(E_{ab,n})
tr(\bar \Gamma^{s^{\p}} \Gamma^{s} )
\ee
$$
P^{ab}_n= (m_{ab,n},0)
\,,\quad\quad
m_{ab,n}:=m_{ab}+E_{ab,n}\,.
$$
We should stress that in the last two equations there is
only small momentum  dynamics. High energies may come from the
external bound state energy insertion.

Some comments are in order. Notice first that for $l\not= 0$ (angular
momentum) the wave function (state) and the energy remain unchanged.
 Notice also that the
s-wave state does not
receive contributions from $l\not=0$ states either. The previous
statement
is true due to the fact that the momentum wave function at zero momentum
for
$l\not=0$ is zero. This means that the new interaction does not couple
$l=0$ states with $l\not=0$ states. This result would change
 if
we kept further terms in the effective lagrangian (see (\ref{lagint}))
 but, of course, these contributions would be
subleading.
\medskip

Let us next calculate the decay constant. In order to do it we
split the current as in the last section. The soft current only gives
a contribution with the low momentum states $g$ in the same way as
the hard current
only gives a contribution with the modified Coulomb bound states.
Thus,
we obtain
\be
\langle 0 \vert J_{\Gamma}^{ba}(0) \vert (ab,g,s); \vec v \rangle=
\langle 0 \vert J_{l,\Gamma}^{ba}(0) \vert (ab,g,s); \vec v \rangle=
-{tr\left(\Gamma^{s} \Gamma \right) \over 2}
f_{ab,g}
\ee
\be
\langle 0 \vert J_{\Gamma}^{ba}(0) \vert (ab,n,s); \vec v \rangle=
\langle 0 \vert J_{h,\Gamma}^{ba}(0) \vert (ab,n,s); \vec v \rangle=
-tr\left(\Gamma^{s} \Gamma \right) \sqrt{N_c \over 2}
{\Phi}_{ab,n} (\vec 0)\,.
\ee
Finally the decay constant reads (changing to relativistic
normalization)
\bea
&&
\label{pdc}
\nonumber
\langle 0 \vert J_{\Gamma}^{ba}(0) \vert (ab,n,s); \vec v \rangle_{F}=
-tr\left(\Gamma^{s} \Gamma \right) \sqrt{m_{ab,n}N_c}
{\Phi}_{ab,n} (\vec 0)
\\
\nonumber
&&
\times
\left\{ 1+
{1 \over 4N_c} {d\P_{ab}(E_{ab,n})\over dE_{ab,n}}
|E_{ab,n} {\tilde \Phi}_{ab,n}(\vec 0) |^2+
{\P_{ab}(E_{ab,n})\over 2N_c} E_{ab,n}
{{\tilde
\Phi}_{ab,n} (\vec 0) \over \Phi_{ab,n} (\vec 0)} \right.
\\
&&\left. \times
\left( 1
+ \sum_{m \not= n} \Phi_{ab,m} (\vec 0) {\tilde
\Phi}_{ab,m}^{*}(\vec 0)
{E_{ab,m} \over E_{ab,n}- E_{ab,m}} \right)
\right\}\,.
\eea
Finally let us obtain the bilinear currents at zero recoil. For
that we need to know
\be
\nonumber
J_{\Gamma}^{bc} (0)=\bar Q^b \Gamma Q^c (0)\,,
\ee
\be
\langle  (ac,N^{\p},s^{\p});
\vec v
\vert J_{\Gamma}^{bc} (0) \vert
 (ab,N,s);
\vec v
\rangle\,.
\ee

In order to deal with them we need to perform the splitting between
large and small momentum.
However this current cannot be in general splitted in two terms. We have
mixing
between large and small momentum. Fortunately, the mixing terms
disappear if both inicial and final
states have the same velocity. This will not longer be true for non-zero
recoil matrix elements. Thus, we obtain
\be
\langle  (ac,n^{\p},s^{\p});
\vec v
\vert J_{\Gamma}^{bc} (0) \vert
 (ab,n,s);
\vec v
\rangle =  -
{tr( \bar \Gamma^{s^{\prime}} \Gamma^{s}
\Gamma)           \over 2}
\int_{\mu} {d^3\vec k\over (2\pi)^3}
\tilde \Phi^{*}_{ac,n^{\prime}}(\vec k)\tilde \Phi_{ab,n}(\vec k)
\ee
\be
\label{meg}
\langle  (ac,g^{\p},s^{\p});
\vec v
\vert J_{\Gamma}^{bc} (0) \vert
 (ab,g,s);
\vec v
\rangle =: -
{tr( \bar \Gamma^{s^{\prime}} \Gamma^{s}
\Gamma)                   \over 2}f^{g^{\p}g}_{ac,ab}
\ee
where we have used
\be
\sum_s
(\Gamma^{s} )_{\alpha_2\alpha_4}
(\bar\Gamma^{s})_{\alpha_1\alpha_3}=-2(p_{+})_{\alpha_2\alpha_3}(p_{-})_{
\alpha_1\alpha_4}\,.
\ee
The remaining possible matrix elements are zero.
Notice that $f^{g^{\p}g}_{ab,ab}=\delta_{g^{\p}g}$ because of baryonic
charge conservation.

The physical matrix element reads (again with relativistic
normalization)
\bea
\label{pme}
\nonumber
&& _F \langle (ac,n^{\prime},s^{\prime});
\vec v
\vert J_{\Gamma}^{bc} (0) \vert
 (ab,n,s);
\vec v
\rangle_F =  -
\sqrt{m_{ab,n}m_{ac,n^{\prime}}}
tr( \bar \Gamma^{s^{\prime}} \Gamma^{s}
\Gamma)
\\
\nonumber
&&
\times \Biggl\{
\int_{\mu} {d^3\vec k\over (2\pi)^3}
\tilde \Phi_{ac,n^{\prime}}^*(\vec k)\tilde \Phi_{ab,n}(\vec k)
\\
\nonumber
&&
\times \Bigl\{ 1+
{1 \over 4N_c} {d\P_{ab}(E_{ab,n})\over dE_{ab,n}}
|E_{ab,n}{\tilde \Phi}_{ab,n}(\vec 0) |^2 +
{1 \over 4N_c} {d\P_{ac}(E_{ac,n^{\p}})\over dE_{ac,n^{\p}}}
|E_{ac,n^{\p}}{\tilde \Phi}_{ac,n^{\p}}(\vec 0)|^2
\Bigr\}
\\
&&
+ {\P_{ac,ab}(E_{ac,n^{\p}},E_{ab,n})\over 2N_c}
E_{ac,n^{\p}} E_{ab,n} \tilde \Phi_{ac,n^{\prime}}^{\ast}(\vec 0) \tilde
\Phi_{ab,n}(\vec 0)
\\
\nonumber
&&
+ {\P_{ac}(E_{ac,n^{\p}})\over 2N_c} E_{ac,n^{\p}}
\sum_{m \not= n^{\prime}} {\tilde
\Phi}_{ac,m}(\vec 0)
{\tilde
\Phi}_{ac,n^{\prime}}^* (\vec 0)
{E_{ac,m}
\over E_{ac,n^{\prime}}- E_{ac,m}} \int_{\mu} {d^3\vec k\over (2\pi)^3}
\tilde \Phi_{ac,m}^{*}(\vec k)\tilde \Phi_{ab,n}(\vec k)
\\
\nonumber
&&
+{ \P_{ab}(E_{ab,n})\over 2N_c} E_{ab,n}
\sum_{m \not= n} {\tilde
\Phi}^*_{ab,m}(\vec 0)
{\tilde
\Phi}_{ab,n} (\vec 0)
{E_{ab,m} \over E_{ab,n}- E_{ab,m}}
\int_{\mu} {d^3\vec k\over (2\pi)^3}
\tilde \Phi_{ac,n^{\prime}}^{*}(\vec k)\tilde \Phi_{ab,m}(\vec k)
\Biggr\}\,,
\eea
where
$$
\int d^4x_1 d^4x_2 e^{iP^{ac}_{n^{\p}} .x_1}e^{-iP^{ab}_n .x_2}
\langle 0 \vert T \left\{ J^{ca}_{l,\bar \Gamma^{s^{\p}}} (x_1)
J^{bc}_{l,\Gamma} (0)
J^{ab}_{l,\Gamma^{s}} (x_2) \right\} \vert 0 \rangle
$$
\be
\label{cor2}
=: \P_{ac,ab}
(E_{ac,n^{\p}} ,E_{ab,n}) tr( \bar \Gamma^{s^{\prime}} \Gamma^{s}
\Gamma)
\ee
 We can easily
check that orthonormality is fulfilled when $b=c$\footnote{This was not
always the case for the result given in \cite{nos1}.}.
 We expect the
last statement to be true since
spin symmetry relates the matrix element with the baryonic
charge when $b=c$.

\be
_F \langle  (ab,n^{\p},s^{\p});
\vec v
\vert J_{\Gamma}^{bb} (0)
 \vert
 (ab,n,s);
\vec v
\rangle_F =
  -
m_{ab,n}
tr( \bar \Gamma^{s^{\prime}} \Gamma^{s}
\Gamma) \delta_{n,n^{\prime}}\,.
\ee
Before finishing this section let us make some remarks.
 Both correlators (\ref{cor1}) and (\ref{cor2}) should be small
quantities for perturbation theory to hold.
This is the case
 if
$\mu$ is
small against the typical momentum in the Coulomb interaction (i.e.
 $\mu a_{ab,n}<<1$, where $a_{ab,n}=n/\mu_{ab}\alpha$ is the Bohr radius).
It constraints the possible applications to the lower energy
levels. On the other hand $\Lambda_{QCD}<<\mu$ should hold so that the
low momentum dynamics is not strongly affected by the cut-off.

%%%%%%%%%%%%%%%%%%%%%%%%%%%%%%%%%%%%%%%%%%%%%%%%%%%%%%%%%%%%%%%%%%%%%%%%%%
\section{Cut-off independence}
\indent

Our results in the last section may look like strongly cut-off
dependent. We
have two sources of cut-off dependence.
 On the one hand we have a cut-off separating small momentum gluons from
large momentum gluons. This cut-off is the responsible for the absence
of Coulomb interation in $L^{\mu}$. It has been mentioned at several
instances but it has never been written down explicitly in the formulas.
This cut-off dependence has been analysed before \cite{nos2}
so we shall ignore it in the following. On the other hand we have the
cut-off separating large and small relative momenta. It plays the role
of an
infrared cut-off in the perturbative Coulomb wave function (large
momentum) and the role of an ultraviolet cut-off
 for the small momentum contributions.
 We prove in this section the cut-off independence to the desired
order ($\mu^3, \Lambda_{QCD}^3$) of this last cut-off. This is crucial
to ensure that our approach respects colour $SU(3)$ gauge symmetry.
It is important to use the same cut-off procedure in
both large and small momentum regions in order to neatly cancel the
cut-off dependence. We
use a hard three momentum cut-off for convenience, as we have done in
the previous sections.

\medskip

First of all, let us study the cut-off dependence in the low
momentum correlators we
found in the last section. Although they are non-perturbative objects we
can always perform a perturbative calculation in order to see how they
depend on the cut-off.

Let us start by (\ref{Lmu}) ( which is formally equal to NRQCD).
For (\ref{cor1}) we obtain at the lowest order (in the CM frame, $\vec
v=0$) \be
\label{pipert}
\P_{ab}(k^0)=-{N_c\mu_{ab}\mu \over \pi^2} \left[ 1-{1 \over
2x}ln\left( {1+x \over 1-x} \right) \right], \quad x={\mu \over
\sqrt{2\mu_{ab}(k^0+i\epsilon)}}\,.
\ee
Let us consider two limits.

In the limit $x>>1$ (i.e. near threshold) it reduces to
\be
\P_{ab}(k^0)\simeq -{N_c\mu_{ab}\mu \over \pi^2} \left[ 1+{i\pi
\over 2x}\theta(k^0) \right] \,
\ee
and no pole appears. In the limit $x<<1$ (\ref{pipert}) reduces to
\be
\label{pihqet}
\P_{ab}(k^0)\simeq {N_c\mu^3 \over 6 \pi^2} {1 \over
(k^0+i\epsilon)}\,.
\ee
This expression is going to be important in the following. We stress
that (\ref{pihqet}) is $\mu_{ab}$ independent, and amounts to drop the
$1/m$ terms in (\ref{Lmu}) which is nothing but the HQET for
quarks and antiquarks. Let us now look for the physical
situation we are interested in. Thus, we take $k^0=E^0_{ab,n}$ and
we obtain $|x|=
{\mu n \over \mu_{ab} C_F \a_s}$, but this is nothing but the
parameter we need to keep small so that the small relative momentum
contributions are subleading, and hence our
expansion makes sense.
In the following,  we always consider that we are in the limit $|x|<<1$.

For (\ref{cor2}) we obtain ($x<<1$)
\be
\label{pi1hqet}
\P_{ac,ab}(k^{\p 0},k^0)\simeq
{N_c\mu^3 \over 6 \pi^2} {1 \over
(k^0+i\epsilon)}
{1 \over
(k^{\p 0}+i\epsilon)}\,.
\ee
At this point we would like to stress that in the limit $x<<1$ the same
perturbative results are found using HQET. This is going to
be determinant in the next section.

\medskip

(\ref{pihqet}) and (\ref{pi1hqet}) make explicit the UV cut-off
dependence coming from the
small relative momentum region. Let us next go on with the IR cut-off
dependence coming from the large relative momentum region.

\medskip

\medskip

Let us then study the cut-off dependence of the wave function (for
simplicity we omit the flavour indices).
 In order to do it we solve the wave equation (\ref{mueq})
perturbatively in $\mu$. $n$ labels continuum or discrete spectrum.
Because of the radial symmetry, we can write (we follow
\cite{Bethe})

\be
\label{phinl}
\tilde \Phi_{n,l,m} (\vec p;\mu) = F_{n,l}(p;\mu)Y_{l,m}({\hat p})
\ee
where $F_{n,l}(p;\mu)$ fulfils
\be
{p^2 \over 2\mu_{ab}} F_{n,l}(p;\mu)-{C_{F} \alpha \over
\pi p} \int_{\mu}q dq F_{n,l}(q;\mu) Q_{l}({p^2+q^2 \over 2pq}) =
E_{n}(\mu) F_{n,l}(p;\mu)
 \quad  p > \mu
\ee
\be
Q_{l}(z)= {1 \over 2} \int^{1}_{-1} dx {P_{l}(x) \over z-x}
\ee
and $P_{l}$ is the Legendre function of the first kind.
We stress that we are interested in $F_{n,l}(p;\mu)$
for $p>\mu$ only,
although in the intermediate steps it is going to be defined over all
$p>0$ values. Now we perform a cut-off parameter expansion and we work
as in usual quantum mechanics
perturbation theory where we demand the corrections to be orthogonal
to the leading result.
\be
\label{Fnl}
F_{n,l}(p;\mu)= \sum_{r=0}^{\infty} F_{n,l}^r(p){\mu^r \over
r!} \quad
E_{n}= \sum_{r=0}^{\infty} E_{n}^r{\mu^r
\over r!} \,.
\ee

We also expand the cut-off in the integral.
\be
\int_{\mu}q dq F^r_{n,l}(q) Q_{l}({p^2+q^2 \over 2pq}) \equiv
h^r_{n,l}(p,\mu) \equiv
\sum_{i=0}^{\infty} h_{n,l}^{r,i}(p){\mu^i \over i!}\,.
\ee

On general grounds we can see that the corrections to the Coulomb
wave function and energy go like $O(\mu^{2l+3})$, therefore,
as expected, we can neglect the $l \not= 0$ states since their
contributions are subleading.

At leading order we obtain the standard Schr\"odinger equation
with a Coulomb potential with no $\mu$ dependence.
Furthermore, for the following terms in perturbation theory we obtain
\be
\label{E1}
E^1_n=E^2_n=0 \,,\quad
{\tilde \Phi}^1_{n}(\vec p)={\tilde \Phi}^2_{n}(\vec p)=0\,.
\ee
Finally to third order we obtain
\be
\label{E3}
E^3_n= -E_n {|\tilde \Phi_{n} (\vec0)|^2 \over \pi^2}\,,
\ee
\be
\label{F3}
{\tilde \Phi}^3_{n}(\vec p)=\sum_{m \not= n}{\tilde \Phi}_{m}(\vec p)
{\tilde \Phi^*_{m}(\vec 0) \tilde \Phi_{n}(\vec 0) \over \pi^2}
{E_m \over E_m - E_n}\,.
\ee
We have not yet normalized the cut-off dependent wave function, as
we can see from

\be
\int_{\mu}{d^3q \over (2\pi)^3}
\tilde \Phi_{n}^{*}(\vec q) \tilde \Phi_{n}(\vec q) \simeq
1 -
|\tilde \Phi_n(\vec 0)|^2
{\mu^3 \over 6 \pi^2}\,. \ee
Therefore, we must change
\be
\label{normcoul}
\tilde \Phi_{n} (\vec p;\mu)
\rightarrow
\tilde \Phi_{n} (\vec p;\mu)
\left( 1 +
|\tilde \Phi_n(\vec 0)|^2
{\mu^3 \over 12 \pi^2} \right) \,. \ee

\medskip

(\ref{E3})-(\ref{normcoul}) provide the explicit IR cut-off
dependence from the large relative momentum region.

\medskip

We have obtained the explicit cut-off dependence to the desired order
$\mu^3$ in both large and small momentum regions. Now we will see they
match
properly, that is, the observables are cut-off independent. In fact what
we will see is that the physical states (\ref{phstate}) themselves are
already
cut-off independent. In this way we prove the cut-off independence for
any observable.

\medskip

Consider first the bound state energy $E_{n}^F$
\be
E_{n}^{F}=E_{n}+\delta E_{n}\,.
\ee
The cut-off dependence of $E_{n}$ is given by (\ref{Fnl}),
(\ref{E1}) and (\ref{E3}), whereas
the cut-off dependence of $\delta E_{n}$ is given by (\ref{deltam})
and (\ref{pihqet}).
One can then easily check that $E_{n}^{F}$ is cut-off independent.

Consider next the state $
 \vert
\overline{(ab,n,s);
\vec v}
\rangle_F
$ in (\ref{n}). Recall that the first and last term on the rhs
belong to the
large relative momentum region whereas the term in the middle belongs to
small relative momentum region. Let us keep apart for a moment the
explicit cut-off separating these two regions in the relative momentum
integrals.
 The remaining cut-off
dependences of the first term
are given by (\ref{phinl}), (\ref{Fnl}), (\ref{E1}) and (\ref{F3}),
while for the last term are given by (\ref{n2}) and (\ref{pihqet}),
 which cancel each other.

It remains the UV cut-off dependence coming from (\ref{n1})
(which
has been already studied in \cite{nos1}) and the explicit IR cut-off
dependence coming from the integral over relative momentum in the
first term of (\ref{n}) (see (\ref{nstate})), which we kept apart for a
while.
Recall that the wave function in the first term of (\ref{n}) is,
 except for the normalization factor (\ref{normcoul}),
the Coulomb wave function
 since we have already cancelled the cut-off dependences
coming from (\ref{F3}).
Let us next calculate (\ref{n1}) perturbatively at lowest order.
 It reads
\bea
\nonumber
&&\vert (ab,n,s);\vec v
\rangle^{(1)}=
{\tilde \Phi_{ab,n} (\vec 0) \over \sqrt{2N_c}}
\int^{\mu} {d^3\vec k
\over (2\pi)^3}
\bar u^{\alpha} (p_1) \Gamma_{s} v^{\b}(p_2)
b^{a \dagger}_{\a,i}(p_1)d^{b \dagger}_{\b,i}(p_2) \vert 0\rangle
\\
&&=
{1 \over \sqrt{2N_c} }
\int^{\mu} {d^3\vec k
\over (2\pi)^3}
\tilde \Phi_{ab,n} (\vec k)
\bar u^{\alpha} (p_1) \Gamma_{s} v^{\b}(p_2)
b^{a \dagger}_{\a,i}(p_1)d^{b \dagger}_{\b,i}(p_2) \vert 0\rangle \,.
\eea
The second equality holds at the order we are working at. Notice finally
that this is nothing but the piece we need to add to the first term of
(\ref{n})
in order to obtain a relative momentum integral independent of the cut-off.
Finally, the cut-off dependence of the normalization in (\ref{normcoul}) and
of (\ref{normstate}) also
cancel each other in (\ref{phstate}) (again taking into account
(\ref{pihqet})).

\medskip

We have thus seen that at the level
of physical states we are able to prove the cut-off independence.
The cut-off independence can also be checked explicitely in the
observables (\ref{deltam}), (\ref{pdc}) and (\ref{pme}).
This demonstrates that the HQET ultraviolet behavior cancels
the NRQCD infrared behavior in Coulomb type bound states, which
garanties that we have performed a proper matching
between large and small relative momentum. This issue has also been
pursued in \cite{nos1,nos2}.

%%%%%%%%%%%%%%%%%%%%%%%%%%%%%%%%%%%%%%%%%%%%%%%%%%%%%%%%%%%%%%%%%%%%%%%%%%%
\section{Evaluation of the low momentum correlators}
\indent

In section 3 we learnt how to parametrize the possible
non-perturbative contributions in the small relative momentum region in
terms of
two low momentum
correlators ((\ref{cor1}) and(\ref{cor2})) with external Coulomb bound
state energy insertions. It is remarkable that these contributions only
exist for s-states. At the beginning of section 4 we also saw that the kinetic
 term, which is suppressed by a mass invers power, can be safely neglected
in the correlators we are interested in, and hence we can use HQET for
quarks and antiquarks to discuss their properties.

The HQET for quarks and antiquarks enjoys a $U(4N_{hf})$ symmetry which
breaks spontaneously down to the
 $U(2N_{hf})\otimes U(2N_{hf})$
 Isgur-Wise
symmetry \cite{joan2}.

Let us first analyse the consequences of the unbroken
 $U(2N_{hf})\otimes U(2N_{hf})$ symmetry. In fact the spin symmetry which
is included in it has already been used in (\ref{cor1}) and
(\ref{cor2}). The flavour symmetry implies the following
\be
f_{ab,g}=f_g \,, \quad f^{g^{\p}g}_{ac,ab}= f^{g^{\p}g}\,.
\ee
Therefore we get
\be
\P_{ab}(E_{ab,n})=
\P_1(E_{ab,n})
\,, \quad
\P_{ac,ab}(E_{ac,n^{\p}},
E_{ab,n})=
\P_{2}(E_{ac,n^{\p}},
E_{ab,n})\,.
\ee
The correlators (\ref{cor1}) and (\ref{cor2}) are thus given in terms
of two unknown
universal (flavour independent) functions $\Pi_1$ and $\Pi_2$.
But if we go further, using flavour number conservation together
with flavour symmetry, we obtain $f^{g^{\p}g}=\delta_{g^{\p}g}$ for any
flavour. From that it follows that if $\Pi_1$ is known for any energy
insertion we can obtain $\Pi_2$. Explicitely they read
\be
\Pi_1(E_{ab,n})=\sum_{g} {|f_g|^2 \over 2}{1\over E_{ab,n}-E_g}\,,\quad
\Pi_2(E_{ac,n^{\p}},E_{ab,n})=\sum_{g} {|f_g|^2 \over 2}
{1\over E_{ac,n^{\p}}-E_g}\,{1\over E_{ab,n}-E_g}\,.
\ee

 These low momentum correlators
can be further specified at least in two situations.

\medskip

i) $E_{ab,n} >> \Lambda_{QCD}\,, \quad
 (m_{Q} \rightarrow \infty \,, \quad \alpha \; small)
$\,,

\medskip

ii) $E_{ab,n} << \Lambda_{QCD} \,, \quad
 (m_{Q} \; large \,, \quad \alpha \rightarrow 0)
$\,.

\medskip

Notice that situation ii) is conceivable if $\alpha$ is very
small since so far we have only assumed that the invers Bohr radius is
much bigger than $\Lambda_{QCD}$ and the energy is suppressed by a
factor $\alpha$ with respect to the former\footnote{In practise we
must remember that $\alpha$ should better be substituted by the running
coupling constant at the quarkonium scale, which is in fact an
implicit function of $m_{Q}$ and $\Lambda_{QCD}$.}
\footnote{In \cite{nos1}, the bound state energy $E_{ab,n}$ was
understood as giving rise to a residual mass for the heavy quark and
antiquark in the Heavy Quark Effective lagrangian, which was latter on
subtracted. That definitively obscures its actual role, which eventually
led to some confusion: in \cite{nos1} the situation ii) was not
allowed whereas the Heavy Quark Hadronic lagrangian was used for
situation i), which is not correct.}.

\medskip

In the situation i) the operator product expansion holds. If we carry
it out for the low momentum correlators we just obtain (\ref{pihqet})
and (\ref{pi1hqet}). Their
cut-off dependence just cancels the cut-off dependence from
the large relative momentum region, as we saw in section 4. Hence, we
conclude that there are no new non-perturbative contributions in this
situation,
 thus confirming
the fact that the VL contributions from the condensate are indeed the
leading non-perturbative effects in
the $m_{Q}\rightarrow \infty$ limit\footnote{This point was
not properly specified in \cite{nos1}.}. This result follows from
the
observation that there is no local gauge invariant object that can be
built out of $D_0$ alone. We have explicitely checked it for lower order
terms.

In the situation ii) we are in the low energy regime of the HQET.
In this regime it is important that
the HQET with quarks
and antiquarks with the same velocity undergoes a spontaneous
symmetry breaking of a $U(4N_{hf})$
symmetry down to the Isgur-Wise symmetry
$U(2N_{hf})\otimes
U(2N_{hf})$,
since the Goldstone modes associated to the broken generators
 dominate the dynamics.
 The Heavy Quark Hadronic effective lagrangian describing
 the Goldstone modes
 was
worked out in \cite{nos1}, where the correlators (\ref{cor1}) and
(\ref{cor2}) were also calculated.
 Using those results we obtain
\be
\label{cor1nos1}
\P_{ab}(k^0)= {
f_{H}^2
 \over 2} {1 \over
(k^0+i\epsilon)}\,,
\ee
\be
\label{cor2nos1}
\P_{ac,ab}(k^{\p 0},k^0)=
{f_{H}^2 \over 2} {1 \over
(k^0+i\epsilon)}
{1 \over
(k^{\p 0}+i\epsilon)}\,,
\ee
\be
{f_{H}^2\over 2}=
{{\bar f_{H}}^2\over 2}
+{\mu^3 N_{c}\over 6\pi^2}
\ee
where $
{\bar f_{H}}^2
$ is cut-off independent.
Notice that in this situation all non-perturbative effects in the
small relative momentum region are parametrized by a single
nonperturbative constant which is spin and flavour
independent\footnote{Notice also that although at
first sight the contributions obtained by substituting (\ref{cor1nos1})
and (\ref{cor2nos1}) in (\ref{deltam}), (\ref{pdc}) and (\ref{pme}) look
like
more important than those from the condensate when $m_{Q}\rightarrow
\infty$, they are actually not so since the smallness of $\alpha$
required in situation ii) mantains the condensate contribution
dominant.
Some statements made in \cite{nos1} implying the opposite must be
corrected.}.
 This is a non-trivial consequence of the $U(4N_{hf})$ symmetry.
The fact that the latter is spontaneously broken down
to
$U(2N_{hf})\otimes
U(2N_{hf})$
allows us to know the Green function behaviour at low energy insertion
with a single nonperturbative constant since no mass term appears in the
pole.
All the spin and flavour dependence is explicitely
known in the observables.

\medskip

However, caution must be taken in the situation ii). This is due to
the fact that, in this situation
the standard evaluation of
non-perturbative contributions
in the large
relative three momentum region coming from (\ref{lagvol})
becomes unreliable. Let us briefly recall the two
approximations involved, namely the multipole expansion and the
adiabatic
approximation. The first one is an expansion in $\Lambda_{QCD}$ over the
invers Bohr radious, which has also been assumed to hold throughout.
 The second one requires the time evolution of the soft
gluon fields to be slow in comparison with the energies involved in the
Coulomb spectrum. This requirement is in fact the opposite of situation
ii). Thus we are in the unfortunate even that when we have an excelent
parametrization of the non-perturbative effects in the small relative
momentum region ((\ref{cor1nos1}) and (\ref{cor2nos1})) we loose control
of them in the large relative momentum region.

\medskip

Nevertheless, we envisage a situation where the parametrization
(\ref{cor1nos1}) and (\ref{cor2nos1})
may be useful. Recall that although the parameter controlling
the
adiabatic approximation and the parameter controlling the expansion in
the hadronic effective lagrangian are both of order $\Lambda_{QCD}$,
they need not be exactly the same. The former was shown to be
$<DFDF>/<FF>$ in \cite{Vol} and let $2\pi {\bar f}_{H}^{2\over 3}$ be
the latter. Suppose then that
$$
E_{ab,n}> \left( {<DFDF> \over <FF>} \right) ^{1\over 2}\,,
$$
\be
\label{<>}
E_{ab,n}< 2\pi{{\bar f}_{H}}^{2\over 3}\,.
\ee
In such a situation it would be reasonable to use both the adiabatic
approximation in the large relative momentum region and the hadronic
effective lagrangian in the small relative momentum region.
Some bottomonium, charmonium, and
presumably $B_{c}$
states may well be considered in the situation (\ref{<>}). However, the
mass of the
 $b$ quark and mainly the mass of the $c$ quark are not large enough to
allow
for a straightforward application of our formalism to phenomenology.
Relativistic and radiative corrections are in general
important and this is also so for the non-perturbative
corrections due to the gluon condensate \cite{Ynd}. All them must be
taken into account.

\medskip

Let us next discuss the expected size of our
contributions.
It is not our aim to present a full-fletched phenomenological analysis
in order to extract
 $\bar f_{H}^2$
from the data, which would definitely
be premature as it should
be clear in the following discussion, but just give reasonable estimates
of the expected
magnitude of its contributions. For simplicity, we will concentrate
on the mass corrections.

\medskip

We start with the bottomonium system where our formalism is expected
to apply for the lowest lying states \cite{Ynd,Pant}. We proceed
as follows. First of all, we fix
 $m_b$ and $a_{bb,1}^{-1}$
 using the
experimental data and the available theoretical results while
ignoring the contribution
 $\delta E_{ab,n}$
 in
(3.17). Then we estimate the size of
 $\delta E_{ab,n}$
by letting
$\bar f_{H}^2$
 run within values of the order of
$\Lambda_{QCD}$. We should keep in mind that although we will take
$\bar f_{H}^2$ positive for definiteness it can also be negative.
We extract $m_b$ and $a_{bb,1}^{-1}$
from
  the
selfconsistency equation $a_{bb,1}(\alpha_s(a_{bb,1}^{-1}))=a_{bb,1}$ ,
 and the
$\Upsilon (1s)$ mass. We use the following equation to fit the latter

\begin{equation}
 m_{\Upsilon (1S)} =2m_{b}+A_2+A_3+A_{VL}
\end{equation}
where
\begin{equation}
 a_{bb,1}^{-1} = {m_b C_f {\tilde \alpha}(a_{bb,1}^{-1}) \over 2}
\end{equation}

$$
A_2 = -2m_b {C_f^2 {\tilde \alpha}^2(a_{bb,1}^{-1}) \over 8}
$$
$$
A_3 = -2m_b {C_f^2 \beta_0 \alpha^2 (a_{bb,1}^{-1}) {\tilde \alpha}
(a_{bb,1}^{-1}) \over 8 \pi }
\left( \ln \left[ {(a_{bb,1}^{-1}) \over m_b C_f {\tilde
\alpha}(a_{bb,1}^{-1})} \right] +1 - \gamma_{E} \right)
$$
\begin{equation}
A_{VL} = m_b {e_{10} \pi \langle \alpha_s G^2 \rangle \over (m_b
C_f {\tilde \alpha}(a_{bb,1}^{-1}))^4}
\end{equation}

We have taken the formulas above, which include relativistic, radiative
and the VL non-perturvative corrections, from \cite{Ynd}
\footnote{However, we have
not taken into account the contributions of order
$O(\alpha^4,\alpha^5)$ given in \cite{Ynd} since the complete
calculation at this order is still lacking.}. We allow for different
values of
$\Lambda_{QCD}$ and give the relative weight of each contribution in the
table.
\medskip

Let us next assume that we are in the situation (5.7).
% and, of
%course,
%$a_o^{-1}>
%\Lambda_{QCD}$ is fulfilled.
As mentioned before, this may well be the case for the
$\Upsilon (1S)$,
$\Upsilon (2S)$,
 $\chi_{b}(1P)$, $J/\psi$ (and $\eta_{c}$) and $B_{c}$
(and $B_{c}^{\ast}$).
 If we let
${\bar f}_H^{2/3}$ run between the values

\begin{equation}
E_{ab,n} <2 \pi{\bar f}_H^{2/3} < a_{ab,n}^{-1}
\end{equation}
we can give an estimate of $\delta E_{ab,n}$.
If we allow ${\bar f}_H^{2/3}$  between $100-150 MeV$,
our results turn out to be quite stable under
values of $\Lambda_{QCD}^{nf=3}$s between $200-300$ MeV.
We obtain

\begin{equation}
-9 MeV < \delta E_{bb,1} < -2 MeV
\end{equation}
where the explicit expresion used for calculating $\delta E_{ab,n}$
reads

\begin{equation}
 \delta E_{ab,n} =
-4\mu_{ab} {16 \pi {\bar f}_H^{2} \over N_c C_f {\tilde \alpha}_s(
a_{ab,n}^{-1}) } \left( {n \over 2\mu_{ab}} \right)^3
\end{equation}
Although the smallness of the result above is discouraging at
first sight, it justifies the procedure used and makes it
selfconsistent.
\medskip

For $n=2$
we obtain

\begin{equation}
-55 MeV < \delta E_{bb,2} < -15 MeV
\end{equation}
Recall that only the s-wave states receive this correction. If the sign
of $\bar f_{H}^2$ was negative, the signs above would be reversed. This
would help to understand the mass difference between the $\chi_{b}(1P)$
and the $\Upsilon (2S)$.

Let us finally give some estimates for
 $\delta E_{cc,1}$ and
 $\delta E_{bc,1}$
 corresponding to the
$J/\Psi$ (and $\eta_{c}$) and
the
$B_c$ (and
$B_c^{\ast}$)
 ground states. We have taken the mass of the charm
$
m_c=1570 MeV
$
as given in reference \cite{Ynd}.
For the $J/\Psi$ we find
$$\delta E_{cc,1} \sim -42 MeV$$
 taking $\Lambda_{QCD}^{nf=3}=300 MeV$, $a_{cc,1}^{-1}
= 848 MeV$  and
${\bar f}_H^{2/3}= 150 MeV$.
For the $B_c$ we find
$$\delta  E_{bc,1} \sim -23 MeV$$
 taking
$\Lambda_{QCD}^{nf=3}=300 MeV$, $a_{bc,1}^{-1} =
1013 MeV$
%(this result follows from the selfconsistency equation)
 and
${\bar f}_H^{2/3}= 150 MeV$.

 The above contributions for the energy shifts are, on the one hand,
small
enough to make us confident that our results are under control and, on
the other hand, large enough to hope for its eventual observation.
However, it is important to realize that
the VL contributions are excedingly big for
$\Upsilon (2S)$, $\chi_{b}(1P)$, $J/\psi$ (and $\eta_{c}$) and $B_{c}$
(and $B_{c}^{\ast}$).
 We suspect that the framework used so far to calculate the VL
contributions in the large relative momentum region is not
appropiated for these states. We believe that in order to
make realistic QCD-based predictions for these states one should devise
a reliable
approximation
in the large
relative momentum region
to deal with the situation (ii) above
, namely invers Born
radius and energy larger and smaller than $\Lambda_{QCD}$ respectively.
Work in this direction is in progress \cite{nos4}.

\bigskip

\section{Conclusions}
%\end{document}
\indent

We are confident that the theoretical framework above is going to be
useful for
an eventual QCD-based formalism attempting to encompase situations where
the Coulomb energy is large (small $n$) and situations where it is small
(large $n$) with respect to $\Lambda_{QCD}$ in heavy quarkonium. Even
more, this formalism could also be useful in order to obtain explicitely
the
perturbative Coulomb corrections to the non-perturbative heavy quarks
bound states (large $n$).

\medskip

Our formalism is clearly inspired by the Wilson renormalization group
approach. We separate the fields into large and small momentum components
by an explicit cut-off,
 and work out what the effective action for the latter is.
However, there is an important point which makes our formalism rather
peculiar: integrating out the large momentum components does not give
rise to local counterterms only. There is non-trivial physics in the
ultraviolet, namely Coulomb type bound states. As far as we know, this
is the first example of a Wilsonian approach where effects due to bound
states have been taken into account.

\medskip

Let us finally summarize the main contributions of this paper.
Elaborating on the ideas first presented in \cite{nos1}, we have
produced a
detailed derivation of the effective theory governing the small relative
momentum degrees of freedom in heavy quarkonium. In
particular this includes an interaction term, which had been overlooked
before, that leads to a few corrections in the observables.
 We have proven the cut-off independence of the formalism.
 We have also
discussed in detail when non-perturbative contributions which cannot
be expressed in terms of local condensates arise, namely when a
description in terms of a Heavy Quark Hadronic Theory is adequate.
Our preliminary estimations suggests that these contributions lead to
energy shifts of a few tens of MeV. Unfortunately, more theoretical
work is necessary to establish them from the data. This is mainly due to
the lack of control on the non-perturbative effects in the large
relative momentum region of most of the systems where our approach
should apply, namely
$\Upsilon (2S)$, $\chi_{b}(1P)$, $J/\psi$ (and $\eta_{c}$) and $B_{c}$
(and $B_{c}^{\ast}$).
 Work in this direction is in progress
\cite{nos4}.

\bigskip

{\bf Acknowledgements}
\medskip

J.S. thanks Prof. G. Veneziano and the CERN theory group for their
hospitality while this work was written up. A.P. acknowledges
a fellowship from CIRIT.
 Financial support from CICYT, contract AEN95-0590 and financial
support from CIRIT, contract GRQ93-1047 is acknowledged.

\bigskip

%%%%%%%%%%%%%%%%%%%%% BIBLIOGRAPHY %%%%%%%%%%%%%%%%%%%%%%%%%%%%%%%%%%%%

\vfill
\eject
%%%%%%%%%%%%%%%%%%%%%%%%%%%%%%%%%%%%%%%%%%%%%%%%%%%%%%%%%%%%%%%%%%%%%%%%%

%
%  Table
%

\begin{table}
\begin{center}
\begin{tabular}{|c|c|c|c|c|c|}  \hline
$\Lambda_{QCD}^{nf=3}(MeV)$
                     & $A_2(MeV)$  & $A_3(MeV)$  & $A_{VL}(MeV)$
                     & $m_b(MeV)$ & $a_{bb,1}^{-1}(MeV)$     \\ \hline
$200 $               & $-314$  & $49$    & $25$
                     & $4850$  & $1234$        \\ \hline
$250 $               & $-376$  & $61$    & $18$
                     & $4879$  & $1354$        \\ \hline
$300 $               & $-440$  & $74$    & $13$
                     & $4906$  & $1468$        \\ \hline
\end{tabular}
\end{center}
\caption{We display A2, A3
and $A_{VL}$ defined in (5.10). The last two columns give our results
for $m_b$ and $a_{bb,1}^{-1}$.}
\end{table}

%
% End table
%

\end{document}